\documentclass[notitlepage]{revtex4-2}

\usepackage[utf8]{inputenc}
\usepackage[dvipsnames]{xcolor}
\usepackage[colorlinks=true,citecolor=RedViolet,urlcolor=BlueGreen,linkcolor=RedViolet]{hyperref}
\usepackage[normalem]{ulem}
\usepackage{url}
\usepackage{graphicx,wrapfig,float,slashed,cancel}
\usepackage{amsmath,amssymb,epsfig,xcolor,stmaryrd}
\usepackage{bm}
\usepackage{enumitem}
\usepackage{hhline,multirow,tabularx} 
\usepackage{graphics}
\usepackage{latexsym}
\usepackage{rotating}
\usepackage{subfigure}
\usepackage{color}
\usepackage{blindtext}
\usepackage{lipsum}
\usepackage{multirow}
\usepackage{cleveref}
\usepackage{orcidlink}

\allowdisplaybreaks
\begin{document}
	
	
	\title{Z boson production in lead-lead collisions: An investigation Utilizing MRW nuclear TMDs}
	
	\author{The CHROMA Collaboration:\\
		S.~Rezaie$^{1}$\orcidlink{0000-0002-1852-1619}}
	\email{somayeh@ipm.ir}	
	
	\author{R.~Kord Valeshabadi$^{1}$\orcidlink{0000-0001-5326-2459}}
	\email{ramin.kord@ipm.ir}
	
	\author{K.~Azizi$^{2,3,1}$\orcidlink{0000-0003-3741-2167}}
	\email{kazem.azizi@ut.ac.ir}
	
	\affiliation{
		$^{1}$School of Particles and Accelerators, Institute for Research in Fundamental Sciences (IPM), P.O. Box 19568-36681, Tehran, Iran\\
		$^{2}$Department of Physics, University of Tehran, North Karegar Avenue, Tehran 14395-547, Iran\\
		$^{3}$Department of Physics, Dogus University, Dudullu-Umraniye, 34775
		Istanbul, Turkiye
	}

    \begin{abstract}
    We investigate $Z$-boson production in Pb--Pb collisions at $\sqrt{s_{NN}}=5.02~\mathrm{TeV}$ within the $k_t$-factorization framework, using four nuclear transverse-momentum-dependent parton distribution (TMD) prescriptions, namely LO-MRW, VO-MRW, MVO-MRW, and MKMRW. The aim of this study is to examine how different ordering conditions and normalization treatments influence the theoretical description of this process. We calculate the rapidity and transverse-momentum distributions using the parton-level KaTie event generator and compare the predictions with CMS measurements and collinear NLO calculations. This comparison allows us to assess the sensitivity of the predicted distributions to the choice of UPDF prescription.
    \end{abstract}
	
	
	\maketitle
	
	\renewcommand{\thefootnote}{\#\arabic{footnote}}
	\setcounter{footnote}{0}
	
	\section {Introduction}
In the usual collinear factorization approach, the transverse momenta of the incoming partons are neglected in the hard-scattering calculation. At Born level, the subprocess $q\bar q\to Z$ therefore produces a $Z$-boson with zero transverse momentum, and nonzero $p_T^Z$ arises from QCD radiation. In $k_t$-factorization, the incoming partons can carry transverse momentum, allowing the Born subprocess itself to contribute at nonzero $p_T^Z$~\cite{Catani:1990eg,Collins:1991ty}. The corresponding parton densities are described by unintegrated parton distribution functions (UPDFs), which enter the calculation together with the off-shell hard-scattering matrix elements.

In Pb--Pb collisions, $Z$-boson production provides a clean probe of the initial state through its leptonic decay channels, since the decay leptons do not undergo strong interactions with the produced medium. The CMS Collaboration has measured the rapidity and transverse-momentum distributions of $Z$ bosons at $\sqrt{s_{NN}}=5.02~\mathrm{TeV}$ for the 0--100\% centrality range~\cite{CMS:2021kvd}. The rapidity distribution is mainly sensitive to the longitudinal momentum fractions and nuclear quark and antiquark densities, whereas the $p_T^Z$ distribution also depends on the transverse momenta of the incoming partons and the recoil from real-emission subprocesses.

Several approaches based on QCD evolution have been developed to describe the transverse-momentum dependence of parton distributions. Among them are the Balitsky--Fadin--Kuraev--Lipatov (BFKL)~\cite{BFKL1,BFKL2,BFKL3,BFKL4} and Catani--Ciafaloni--Fiorani--Marchesini (CCFM)~\cite{CCFM1,CCFM2,CCFM3,CCFM4} evolution equations. The BFKL evolution equation resums logarithms of $1/x$ at high energies, while CCFM accounts for gluon coherence through angular ordering~\cite{KT}. UPDFs can also be obtained from the Dokshitzer--Gribov--Lipatov--Altarelli--Parisi (DGLAP)~\cite{DGLAP1,DGLAP2,DGLAP3} evolution equations. One example is the Parton Branching (PB) method~\cite{PB1,PB2}, which has been applied to Drell--Yan and $Z$-boson production in proton--proton and proton--lead collisions~\cite{BermudezMartinez:2019anj,hanes}. Another widely used DGLAP-based approach is the Kimber--Martin--Ryskin (KMR)~\cite{KMR} and Martin--Ryskin--Watt (MRW)~\cite{MRW} family of prescriptions, in which the transverse momentum is generated at the last step of evolution. These prescriptions have been used in calculations of Drell--Yan and $Z$-boson production at hadron colliders~\cite{Watt:2003vf,Lipatov:2011sd,kord_validity,D,SomZ}, and MRW-type nuclear UPDFs have also been applied to $Z$-boson production in proton--lead collisions~\cite{SomzZz}. The nuclear UPDFs examined in this work belong to the same family.

Although the KMR/MRW prescriptions are based on DGLAP evolution, they differ in the treatment of the last parton splitting and in the normalization of the resulting UPDFs. These differences can affect the transverse-momentum dependence of the UPDFs and the calculated cross sections, even when the same collinear PDFs are used.

The aim of this study is to examine how these differences affect $Z$-boson production in Pb--Pb collisions at $\sqrt{s_{NN}}=5.02~\mathrm{TeV}$~\cite{CMS:2021kvd}. We calculate the rapidity and transverse-momentum distributions using LO-MRW, VO-MRW, MVO-MRW, and MKMRW nuclear UPDFs obtained from the nCTEQ15 nuclear PDFs~\cite{Kovarik:2015cma}. The Born and real-emission subprocesses are evaluated with the parton-level KaTie event generator~\cite{KATIE}, and the predictions are compared with CMS measurements and the collinear NLO prediction reported in Ref.~\cite{CMS:2021kvd}.

The remainder of the paper is organized as follows. Section \ref{sec:kt-factorization} introduces the $k_T$-factorization framework, while Section \ref{sec:3} describes the nuclear UPDF prescriptions. Section \ref{sec:results-discussion} discusses the
numerical implementation and presents the results,
including the separate contributions of the Born and
real-emission subprocesses, the behavior of the
different UPDFs, their predictions for $Z$-boson
production, and comparisons with experimental data. Finally, conclusions are presented in Section \ref{sec:conclusion}.
	\section{$k_t$-factorization framework}
	\label{sec:kt-factorization}
	
	In the $k_t$-factorization framework, the transverse momenta of the incoming partons are included explicitly in both the parton distributions and the hard-scattering matrix elements. The momenta of the incoming partons can be written schematically as
	\begin{equation}
		k_i=x_iP_i+k_{it},
		\qquad i=1,2,
	\end{equation}
	where $P_i$ is the momentum of the parent hadron, $x_i$ is the longitudinal momentum fraction carried by the parton, and $k_{it}$ is its transverse momentum. The incoming partons are therefore generally off shell, with $k_i^2\simeq -k_{it}^2$.
	
	For the production of a dilepton pair through a $Z$ boson in Pb--Pb collisions, the hadronic cross section is obtained by convoluting the nuclear UPDFs with the corresponding off-shell partonic cross sections. Hence, cross section can be written schematically as
	\begin{align}
		\label{eq:kt-cs}
		\sigma_{PbPb}
		={}&
		\sum_{a,b}
		\int \frac{dx_1}{x_1}
		\frac{dx_2}{x_2}
		   dk_{1t}^2 dk_{2t}^2
		F_a^{\mathrm{Pb}}(x_1,k_{1t}^2,\mu^2)
		F_b^{\mathrm{Pb}}(x_2,k_{2t}^2,\mu^2)
		\hat{\sigma}^{*}_{ab},
	\end{align}
	where the sum runs over the partonic channels included in the calculation. In the present analysis, we consider $q\bar q\rightarrow Z \rightarrow \ell^+\ell^-$,
	together with the real-emission subprocesses $qg \rightarrow Z+q \rightarrow \ell^+\ell^-+q$, and $q\bar q \rightarrow Z+g \rightarrow \ell^+\ell^-+g$, including the corresponding crossed initial-state configurations.
	
	Here, $F_{a(b)}^{\mathrm{Pb}}(x,k_t^2,\mu^2)$ denotes the nuclear UPDF of parton ${a(b)}$, while $\hat{\sigma}^{*}_{ab}$ represents the corresponding off-shell partonic cross section. The variables $x_1$ and $x_2$ are the longitudinal momentum fractions of the incoming partons, $k_{1t}$ and $k_{2t}$ are their transverse momenta, and $\phi_1$ and $\phi_2$ denote their azimuthal angles.
	
	The main difference from the collinear framework is that the initial-state transverse momenta enter directly into both the event kinematics and the hard-scattering matrix elements. Therefore, the transverse-momentum dependence of the incoming nuclear partons can affect the kinematic distributions of the produced dilepton system and of the accompanying parton in the real-emission channels.
	
	For the numerical calculation, we employ the \textsc{KaTie}~\cite{KATIE} parton-level event generator to evaluate the off-shell matrix elements and perform the phase-space integration. The nuclear UPDFs are evaluated through \textsc{PDFxUPDFLib}, which is interfaced with \textsc{KaTie} and provides the required UPDFs during the calculation. The nuclear UPDFs used in this work are constructed according to the MRW-type prescriptions described in the following section, namely LO-MRW, VO-MRW, normalization-matched VO-MRW, and modified KMRW(MKMRW).
	\section {The nuclear UPDFs }	
	\label{sec:3}
	\subsection{LO-MRW and VO-MRW UPDFs}
	\label{LO-NLO-MRWSec}
	
	The LO-MRW and VO-MRW prescriptions construct UPDFs from collinear parton distributions evolved according to the DGLAP equations. In this section, we summarize the two prescriptions as used to obtain nuclear quark and gluon UPDFs. Further details of the MRW formalism can be found in Refs.~\cite{MRW,Valeshabadi:2021smo,SomZ,SomzZz}.
	
	In the MRW approach, the parent parton is treated as collinear up to the last
	evolution step. At this step, a parton $b$ described by
	$f_b(x/z,p^2)$ undergoes a splitting governed by
	$\alpha_s(p^2)P_{ab}(z)/(2\pi)$, generating the transverse momentum $k_t$
	of parton $a$. Subsequent evolution up to the factorization scale $\mu$
	without further resolvable emission is described by the Sudakov form factor
	$T_a(p^2,\mu^2)$. The two prescriptions use different scales for the last evolution step. LO-MRW uses $p^2=k_t^2$, whereas VO-MRW uses the parton virtuality $k^2=k_t^2/(1-z)$.
	
	For $k_t^2\geq\mu_0^2$, the LO-MRW distribution is
	\begin{equation}
		\label{eq:LOMRW}
		F_a^{\mathrm{LO-MRW}}(x,k_t^2,\mu^2)
		=
		\frac{1}{k_t^2}
		\frac{\alpha_s(k_t^2)}{2\pi}
		T_a^{\mathrm{LO-MRW}}(k_t^2,\mu^2)
		\sum_{b=q,g}\int_x^1 dz\,
		P_{ab}(z)
		f_b\!\left(\frac{x}{z},k_t^2\right)
		\Theta^{\delta^{ab}}(z_{\max}-z),
	\end{equation}
	where $f_b$ denotes a momentum-weighted collinear parton distribution.
	For $k_t^2\leq\mu^2$, the corresponding Sudakov form factor is
	\begin{equation}
		\label{eq:LOMRWSud}
		T_a^{\mathrm{LO-MRW}}(k_t^2,\mu^2)
		=
		\exp\left[
		-\int_{k_t^2}^{\mu^2}\frac{d\kappa^2}{\kappa_t^2}
		\frac{\alpha_s(\kappa_t^2)}{2\pi}
		\sum_{b=q,g}\int_0^1 d\xi\,
		\xi P_{ba}(\xi)
		\Theta^{\delta^{ab}}\!\bigl(\xi_{\max}(\kappa)-\xi\bigr)
		\right].
	\end{equation}
	For $k_t^2>\mu^2$, we set
	\begin{equation}
		\label{eq:Sud}
		T_a^{\mathrm{LO-MRW}}(k_t^2,\mu^2)=1.
	\end{equation}
	
	The collinear input is defined only above its starting scale $\mu_0$. Following Ref.~\cite{MRW}, we specify the LO-MRW distribution below thisn scale using the normalization condition
	\begin{equation}
		\label{eq:normalMRW}
		f_a(x,\mu^2)
		=
		\int_0^{\mu^2} dk_t^2\,
		F_a^{\mathrm{LO-MRW}}(x,k_t^2,\mu^2).
	\end{equation}
	A constant distribution with respect to $k_t^2$ in the region $k_t^2\leq\mu_0^2$ then gives
	\begin{equation}
		\label{LO-MRW-lowkt}
		F_a^{\mathrm{LO-MRW}}(x,k_t^2\leq\mu_0^2,\mu^2)
		= \frac{1}{\mu_0^2}
		f_a(x,\mu_0^2)
			T_a^{\mathrm{LO-MRW}}(\mu_0^2,\mu^2).
	\end{equation}
	
	The diagonal splitting kernels become singular in the soft-gluon limit. To regulate this singularity, theta functions are introduced in the real-emission term and the Sudakov factor, as shown in Eq.~\eqref{eq:LOMRW} and \eqref{eq:LOMRWSud}. The restriction can be imposed through strong ordering or angular ordering~\cite{MRW,KordValeshabadi:2026unequal}. We use angular ordering, which gives
	\begin{equation}
		\label{eq:AOCutoff}
		z_{\max}=\frac{\mu}{\mu+k_t},
		\qquad
		\xi_{\max}(\kappa_t)=\frac{\mu}{\mu+\kappa_t}.
	\end{equation}
	With this choice, the LO-MRW distribution can receive contributions from $k_t>\mu$.
	It should also be noted that Eq.~\eqref{eq:LOMRW} gives the integral form of the LO-MRW distribution. The prescription is also often written in differential form as~\cite{MRW}
	\begin{equation}
		\label{eq:LOMRWdiff}
		F_a^{\mathrm{LO-MRW,diff}}(x,k_t^2,\mu^2)
		=
		\frac{\partial}{\partial k_t^2}
		\left[
		T_a^{\mathrm{LO-MRW}}(k_t^2,\mu^2)
		f_a(x,k_t^2)
		\right].
	\end{equation}
When ordinary collinear PDFs are used, the differential and integral forms of MRW can give different results. In particular, the differential form can become negative and develop a discontinuity at $k_t=\mu$ \cite{Valeshabadi:2021smo}. Earlier studies proposed remedies for this discrepancy~\cite{Golec-Biernat:2018hqo,Guiot:2019vsm}. While Ref.~\cite{Valeshabadi:2021smo} showed that, for the LO non-singlet distribution, the two forms agree over the full $k_t$ range when cutoff-dependent PDFs are used together with a correction to the differential expression above $k_t=\mu$. Since we use ordinary collinear nuclear PDFs here, we calculate the UPDFs directly from the integral form in Eq.~\eqref{eq:LOMRW}.
	
	In the VO-MRW prescription, the scale of the last splitting is instead
	identified with the virtuality $k^2=k_t^2/(1-z)$. The distribution becomes
	\begin{equation}
		\label{eq:VOMRW}
		F_a^{\mathrm{VO-MRW}}(x,k_t^2,\mu^2)
		=
		\frac{1}{k_t^2}
		\sum_{b=q,g}\int_x^1 dz\,
		\frac{\alpha_s(k^2)}{2\pi}
		T_a^{\mathrm{VO-MRW}}(k^2,\mu^2)
		P_{ab}(z)
		f_b\!\left(\frac{x}{z},k^2\right)
		\Theta^{\delta^{ab}}(z_{\max}-z)
		\Theta(\mu^2-k^2)
		\Theta(k^2-\mu_0^2).
	\end{equation}
	The factor $\Theta(\mu^2-k^2)$ imposes virtuality ordering constraint by requiring the scale of the last splitting to remain below $\mu^2$. Since $k^2=k_t^2/(1-z)$, this also limits the transverse momentum to $k_t^2<(1-z)\mu^2$, and hence $k_t<\mu$. The factor	$\Theta(k^2-\mu_0^2)$ keeps the collinear input above its starting scale. Additionally, the Sudakov factor describes evolution from $k^2$ to $\mu^2$ without further resolvable emission.
	
	\subsection{Normalization-matched VO-MRW UPDFs}
	\label{MVOMRWSec}
	
	The VO-MRW distribution~\cite{MRW} in Eq.~\eqref{eq:VOMRW} generates
	transverse-momentum dependence through the last parton splitting.
	Its integral over $k_t^2$, however, does not necessarily reproduce
	the input collinear distribution because the ordering constraints
	restrict the available phase space. To satisfy the collinear
	normalization condition while preserving the $k_t$ dependence
	generated by VO-MRW, we use the normalization-matched prescription
	introduced in Ref.~\cite{KordValeshabadi:2026unequal}.
	
	The virtuality-ordering condition sets the upper limit of the
	transverse-momentum integral. Since $z\geq x$ and
	$k^2=k_t^2/(1-z)<\mu^2$, the largest allowed transverse momentum is
	\begin{equation}
		\label{eq:MVOktmax}
		k_{t,\max}^2(x,\mu^2)=\mu^2(1-x).
	\end{equation}
	We first integrate the VO-MRW distribution over this range,
	\begin{equation}
		\label{eq:MVOnorm}
		\mathcal{N}_a^{\mathrm{VO-MRW}}(x,\mu^2)
		=
		\int_{0}^{k_{t,\max}^2(x,\mu^2)}
		dk_t^2\,
		F_a^{\mathrm{VO-MRW}}(x,k_t^2,\mu^2).
	\end{equation}
	Therefore, in order to satisfy the Eq.~\eqref{eq:normalMRW}, the matched distribution is then defined as
	\begin{equation}
		\label{eq:MVOMRW}
		F_a^{\mathrm{MVO-MRW}}(x,k_t^2,\mu^2)
		=
		\frac{f_a(x,\mu^2)}
		{\mathcal{N}_a^{\mathrm{VO-MRW}}(x,\mu^2)}
		F_a^{\mathrm{VO-MRW}}(x,k_t^2,\mu^2).
	\end{equation}
	Because the factor multiplying $F_a^{\mathrm{VO-MRW}}$ is independent
	of $k_t$, it changes the normalization without changing the shape
	of the distribution at fixed $x$ and $\mu^2$.
	\subsection{MKMRW UPDFs}
	\label{MKMRWSec}
	As discussed in Sec.~\ref{LO-NLO-MRWSec}, the MRW distribution with
	angular ordering can extend beyond $k_t=\mu$. However, the
	normalization condition in Eq.~\eqref{eq:normalMRW} integrates only
	up to $k_t^2=\mu^2$ and does not include this part of the
	distribution. Motivated by this limitation and the disagreement
	between the differential and integral forms discussed above,
	Ref.~\cite{Guiot:2022psv} introduced the modified KMRW prescription.
	Its normalization condition covers the full transverse-momentum range
	and is given by
	\begin{equation}
		\label{eq:MKMRWnormal}
		f_a(x,\mu^2)
		=
		\int_0^\infty dk_t^2\,
		F_a^{\mathrm{MKMRW}}(x,k_t^2,\mu^2).
	\end{equation}
	
	In this approach the collinear distribution is evaluated at the factorization scale
	$\mu^2$, while the transverse-momentum dependence is introduced through
	the modified Sudakov form factor,
	\begin{equation}
		\label{eq:MKMRWderivative}
		F_a^{\mathrm{MKMRW}}(x,k_t^2,\mu^2)
		=
		\frac{\partial}{\partial k_t^2}
		\left[
		T_a^{\mathrm{MKMRW}}(k_t^2,\mu^2)\,
		f_a(x,\mu^2)
		\right].
	\end{equation}
	The modified Sudakov form factor is defined as:
	\begin{equation}
		\label{eq:MKMRWSud}
		T_a^{\mathrm{MKMRW}}(k_t^2,\mu^2)
		=
		\exp\left[
		-\int_{k_t^2}^{\infty}
		\frac{d\kappa_t^2}{\kappa_t^2}
		\frac{\alpha_s(\kappa_t^2)}{2\pi}
		\sum_b
		\int_0^{\xi_{\max}(\kappa_t)}
		d\xi\,\xi P^R_{ba}(\xi)
		\right],
	\end{equation}
	where $P^R_{ba}$ is the real splitting kernel and
	$\xi_{\max}$ is the same as Eq. \eqref{eq:AOCutoff}. However, unlike the LO-MRW expression, the AO cutoff is applied to every splitting channel in the sum over $b$.
	
	Since $f_a(x,\mu^2)$ does not depend on $k_t^2$, differentiating Eq.~\eqref{eq:MKMRWSud} gives
	\begin{equation}
		\label{eq:MKMRWexplicit}
		F_a^{\mathrm{MKMRW}}(x,k_t^2,\mu^2)
		=
		\frac{f_a(x,\mu^2)}{k_t^2}
		\frac{\alpha_s(k_t^2)}{2\pi}
		T_a^{\mathrm{MKMRW}}(k_t^2,\mu^2)
		\sum_b
		\int_0^{\xi_{\max}(k_t)}
		d\xi\,\xi P^R_{a\to b}(\xi).
	\end{equation}
	The Sudakov factor tends to zero as $k_t^2\to0$ and to unity as $k_t^2\to\infty$. These limits ensure that integrating Eq.~\eqref{eq:MKMRWderivative} gives the normalization condition in Eq.~\eqref{eq:MKMRWnormal}. Additionally, unlike the LO-MRW apporach which uses the low-$k_t$ expression, see Eq.~\eqref{LO-MRW-lowkt}, in this approach similar to VO-MRW method, no separate prescription is needed for the region $k_t<\mu_0$.

\section{Results and discussion}
\label{sec:results-discussion}

We present predictions for $Z$ boson production in Pb--Pb collisions at $\sqrt{s_{NN}}=5.02~\mathrm{TeV}$ using the LO-MRW, VO-MRW, normalization-matched VO-MRW (MVO-MRW), and MKMRW nuclear UPDFs. All four models use the nCTEQ15FullNuc\_208\_82 PDF set as collinear input~\cite{Kovarik:2015cma}. The calculations include the off-shell $q\bar q\to\ell^+\ell^-$, $q\bar q\to\ell^+\ell^-+g$, and $qg\to\ell^+\ell^-+q$ subprocesses. The predictions are compared with the CMS measurement~\cite{CMS:2021kvd} for $60<m_{\ell\ell}<120~\mathrm{GeV}$ and $|y_Z|<2.1$. For comparison, we also include the collinear NLO prediction obtained with MG5\_aMC@NLO using CT14 NLO proton PDFs supplemented by EPPS16 NLO nuclear modifications denoted by MG5+EPPS16, as reported by CMS~\cite{CMS:2021kvd}. The central factorization and renormalization scales are set to the dilepton transverse mass, $\mu_F=\mu_R=\sqrt{m_{\ell\ell}^2+(p_T^Z)^2}$, and the scale uncertainty bands are obtained by varying both scales together by factors of $1/2$ and $2$.

It should be noted that in order to reduce potential double counting between emissions incorporated into the UPDFs and explicit real-emission contributions, we follow the phase-space separation prescription proposed by the Ref.~\cite{Maciula:2019izq} and employed in our previous diphoton study~\cite{Valeshabadi:2025hhr}. In this method for the Born subprocess, the transverse momenta of the incoming partons are restricted to $k_{T,i}\leq\mu_F$, whereas for $q\bar q\to\ell^+\ell^-+g$, the emitted gluon is required to satisfy $p_{T,g}>\max(k_{T,1},k_{T,2})$.

\begin{figure*}[tbp]
	\centering
	
	\begin{minipage}[t]{0.43\textwidth}
		\centering
		\textbf{(a)}\\[-1mm]
		\includegraphics[width=\linewidth]{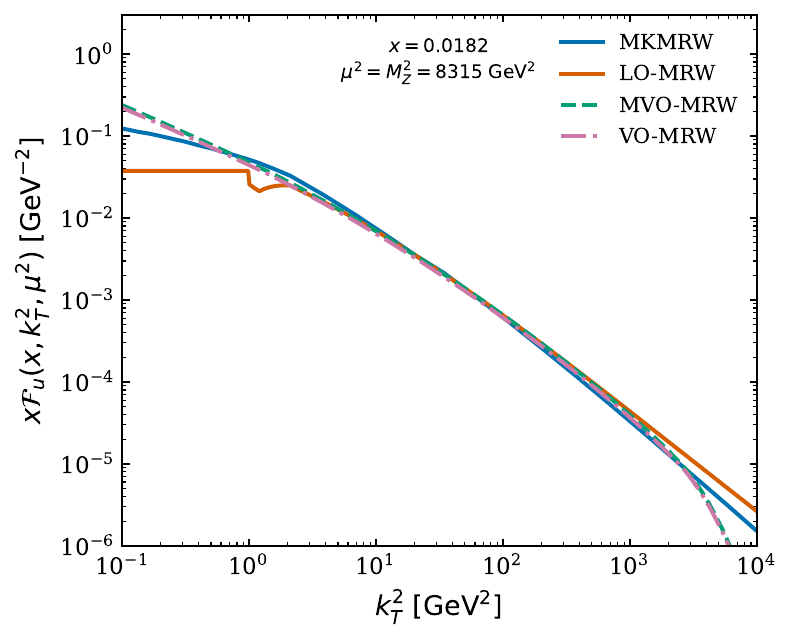}
	\end{minipage}
	\hspace{0.03\textwidth}
	\begin{minipage}[t]{0.43\textwidth}
		\centering
		\textbf{(b)}\\[-1mm]
		\includegraphics[width=\linewidth]{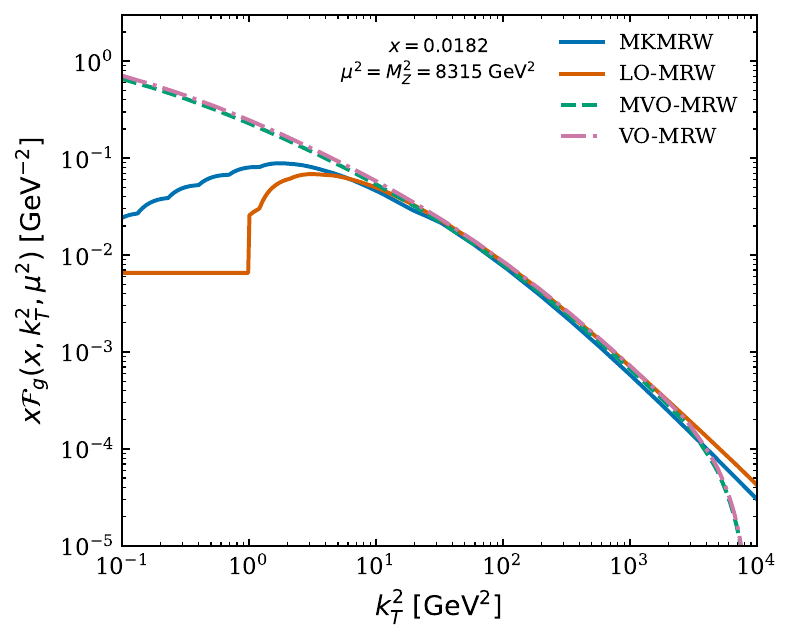}
	\end{minipage}
	
	\caption{
		Nuclear UPDFs obtained with MKMRW, LO-MRW, MVO-MRW, and
		VO-MRW at $x=0.0182$ and $\mu^2=M_Z^2$:
		(a) up-quark UPDF and (b) gluon UPDF.
		The plotted momentum-weighted densities are
		$x\mathcal{F}_u=F_u$ and $x\mathcal{F}_g=F_g$,
		respectively.
	}
	\label{fig:results-updf}
\end{figure*}

To gain insight into how the different UPDF models may affect the differential cross sections studied in this work, we first examine the transverse-momentum dependence of the nuclear UPDFs at $x=0.0182$. The up-quark distributions are shown in the left panel of Fig.~\ref{fig:results-updf}. LO-MRW is constant below $k_t^2=\mu_0^2=1~\mathrm{GeV}^2$, as specified by Eq.~\eqref{LO-MRW-lowkt}. In contrast, the VO-MRW, MVO-MRW, and MKMRW models can be extended directly into the region $k_t<\mu_0$, without requiring the separate low-$k_t$ prescription used in LO-MRW. As shown in the figure, MKMRW can exceed the LO-MRW UPDF in the $k_t<\mu_0$ region, while still remaining below the virtuality-ordered distributions toward the lower end of the $k_t$ range. At intermediate transverse momenta, however, the four predictions become much closer to one another.

The differences become pronounced again toward the upper end of the transverse-momentum range. The virtuality-ordering condition $k^2=k_t^2/(1-z)<\mu^2$, together with $z\geq x$, restricts the VO-MRW and MVO-MRW distributions to $k_t^2<(1-x)\mu^2$. Consequently, both distributions are strongly suppressed as this boundary is approached. In contrast, LO-MRW and MKMRW extend beyond this limit, with LO-MRW giving the larger up-quark distribution in the high-$k_t$ region. This behavior can be attributed mainly to two features of the MKMRW model. One is the additional cutoff imposed on the quark-emission terms, while the other is the stronger suppression from the Sudakov form factor, whose integration range extends to infinity. Additionally, MVO-MRW due to matching factor gives slightly different distribution compared to the VO-MRW model throughout this region.

The gluon distributions in the right panel of the Fig.~\ref{fig:results-updf} show a stronger separation at small $k_t^2$. VO-MRW and MVO-MRW are substantially larger than LO-MRW and MKMRW in the lowest displayed region. MKMRW develops a broad maximum at a few $\mathrm{GeV}^2$, while LO-MRW gives similar constant distribution in the $k_t^2 < \mu_0^2$ region. The distributions become closer at intermediate $k_t^2$, before the virtuality-ordering constraint suppresses VO-MRW and MVO-MRW near their upper boundary. Similar to the up-quark case, the matched gluon distribution has slightly different distribution than the unmatched one.

\begin{figure*}[tbp]
	\centering
	
	\begin{minipage}[t]{0.43\textwidth}
		\centering
		\includegraphics[width=\linewidth]{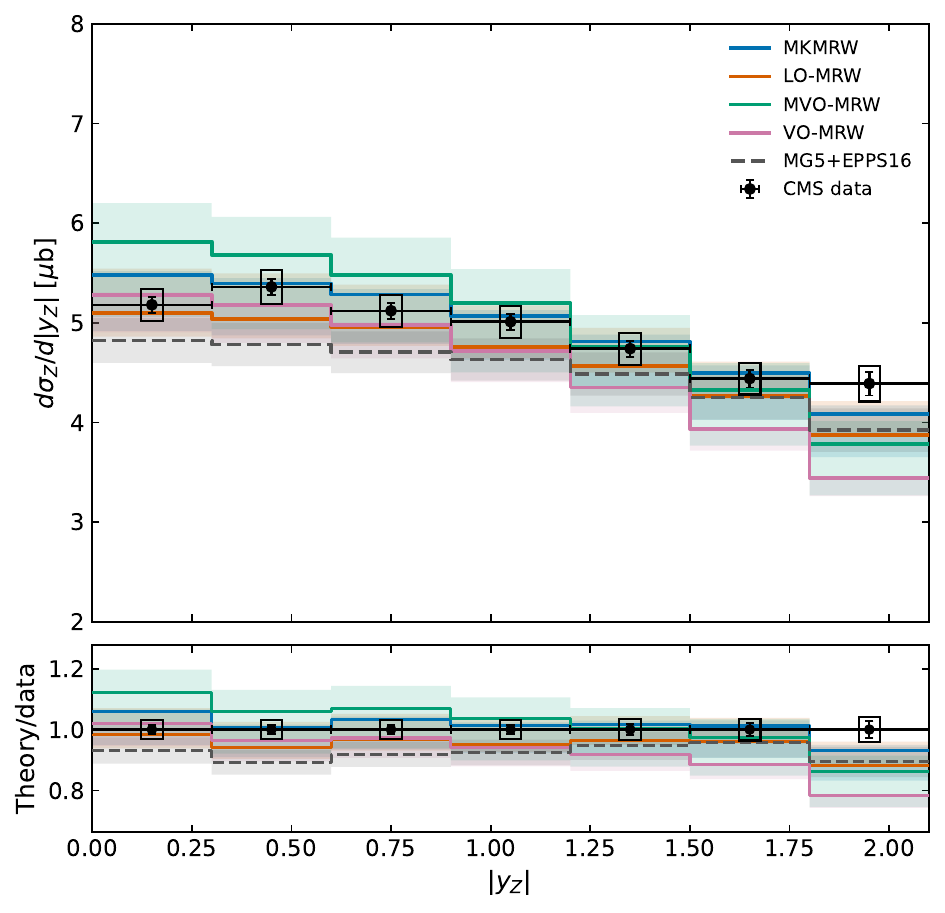}
		
		\vspace{1mm}
		\textbf{(a)}
	\end{minipage}
	\hspace{0.04\textwidth}
	\begin{minipage}[t]{0.43\textwidth}
		\centering
		\includegraphics[width=\linewidth]{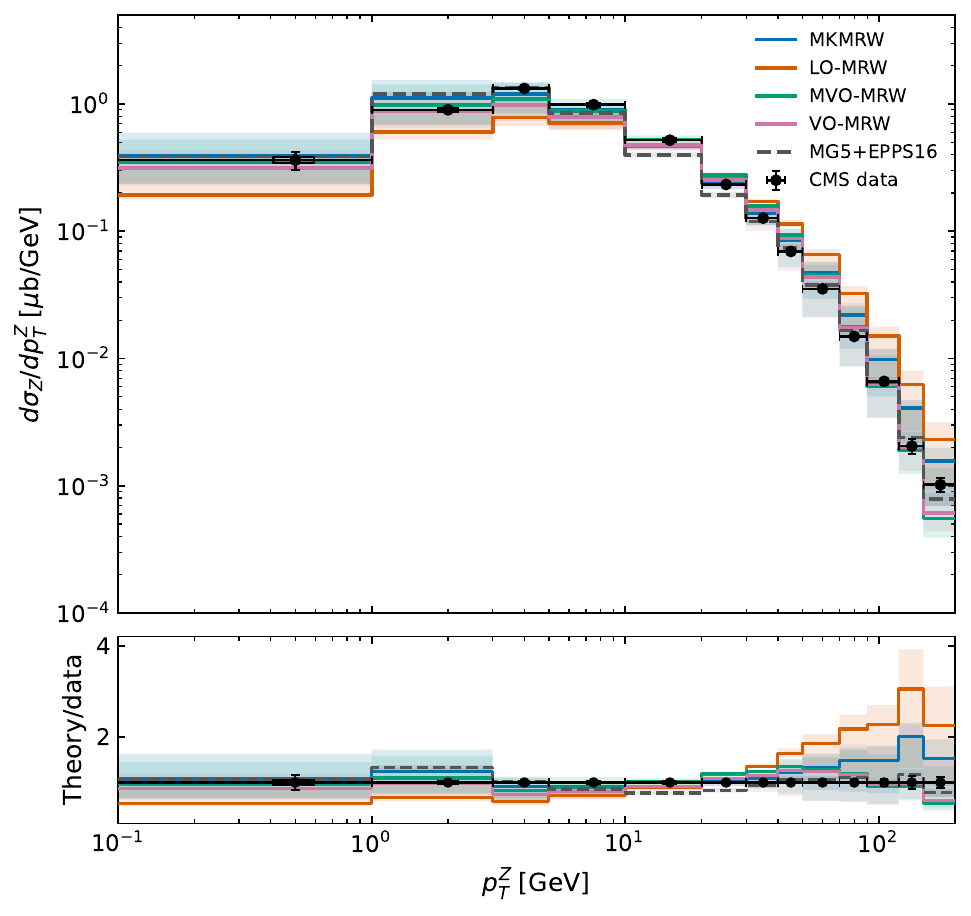}
		
		\vspace{1mm}
		\textbf{(b)}
	\end{minipage}
	
	\caption{
Differential cross sections for $Z$ boson production in Pb--Pb collisions at $\sqrt{s_{NN}}=5.02~\mathrm{TeV}$ as functions of (a) rapidity, for $60<m_{\ell\ell}<120~\mathrm{GeV}$, and (b) transverse momentum, with the additional requirement $|y_Z|<2.1$. The LO-MRW, VO-MRW, MVO-MRW, and MKMRW predictions are compared with CMS measurements and the collinear NLO MG5+EPPS16 calculation. The shaded bands for the UPDF-based predictions correspond to simultaneous variations of the factorization and renormalization scales by factors of $1/2$ and $2$, while the collinear NLO uncertainty band includes PDF and QCD-scale uncertainties combined in quadrature. CMS statistical uncertainties are shown by error bars and systematic uncertainties by open boxes. The lower panels show the corresponding theory/data ratios.
}
	\label{fig:results-cross-sections}
\end{figure*}

The left panel of Fig.~\ref{fig:results-cross-sections} compares the predicted rapidity distributions with CMS data. All four UPDF scenarios, as well as the collinear MG5+EPPS16 calculation, reproduce the overall decrease of the cross section toward larger $|y_Z|$, although differences in normalization and rapidity dependence are observed. Among the UPDF prescriptions, MKMRW generally agrees better with the measured central values than the other UPDF models. It slightly overestimates the cross section at low rapidities, agrees well with the data in the intermediate region, and tends to underestimate the measurements toward forward rapidities. The collinear MG5+EPPS16 prediction also reproduces the decreasing trend but generally lies below the experimental central values. Compared with this collinear reference, MKMRW provides a closer description of the measured rapidity distribution over most of the considered range.

The LO-MRW prediction generally underestimates the measured cross section, with the discrepancy becoming more pronounced toward forward rapidities. The VO-MRW central prediction agrees reasonably well with the data in the lowest rapidity bin but decreases more rapidly than the measured distribution, leading to a noticeable deficit in the forward region. Nevertheless, its scale-uncertainty band remains compatible with the measurements in several low- and intermediate-rapidity bins. Normalization matching shifts the VO-MRW prediction upward, resulting in an overestimation at low rapidities while improving agreement in parts of the intermediate and forward regions. However, the matched prediction still falls below the data in the highest rapidity bin. These results indicate that imposing the collinear normalization condition does not necessarily improve the description of the rapidity distribution uniformly across the full kinematic range.

The transverse-momentum distributions shown in the right panel of Fig.~\ref{fig:results-cross-sections} show larger differences between the models than those observed in the rapidity distributions. One can observe from this figure that LO-MRW underestimates the data at low $p_T^Z$ but gives a harder spectrum, increasingly overestimating the measurements at higher transverse momenta.

It can also be seen from this figure that MKMRW gives a better description of the data at low $p_T^Z$ than LO-MRW, although some differences remain around the peak of the distribution. At higher transverse momenta, however, it also tends to overestimate the measurements. The VO-MRW and MVO-MRW predictions have softer high-$p_T^Z$ tails and agree more closely with the data in this region, except for the last bin, where both underestimate the measured cross section. This behavior is consistent with our previous study of $Z$-boson production in proton--lead collisions~\cite{SomzZz}, where LO-MRW also overestimated the data at high transverse momenta, while the more restrictive NLO-MRW prescription gave better agreement in this region. Normalization matching improves the VO-MRW prediction around the peak while leaving the overall shape largely unchanged. For comparison, the collinear MG5+EPPS16 calculation reproduces the general trend of the measured spectrum, although some discrepancies appear at low and intermediate $p_T^Z$. However, at higher transverse momenta, its prediction remains relatively close to the data, apart from the last bin.

Additionally, the softer tails of VO-MRW and MVO-MRW are consistent with the more restrictive transverse-momentum range imposed by virtuality ordering. However, the transverse momentum of the $Z$ boson depends on both incoming partons and, in the real-emission subprocesses, on the recoil of the outgoing parton. Moreover, the factorization scale increases with $p_T^Z$, changing the allowed transverse-momentum range. Therefore, the differences between the predicted spectra depend not only on the UPDFs but also on the contributions of the different partonic subprocesses.

\begin{figure*}[tbp]
	\centering
	\includegraphics[width=0.98\textwidth]{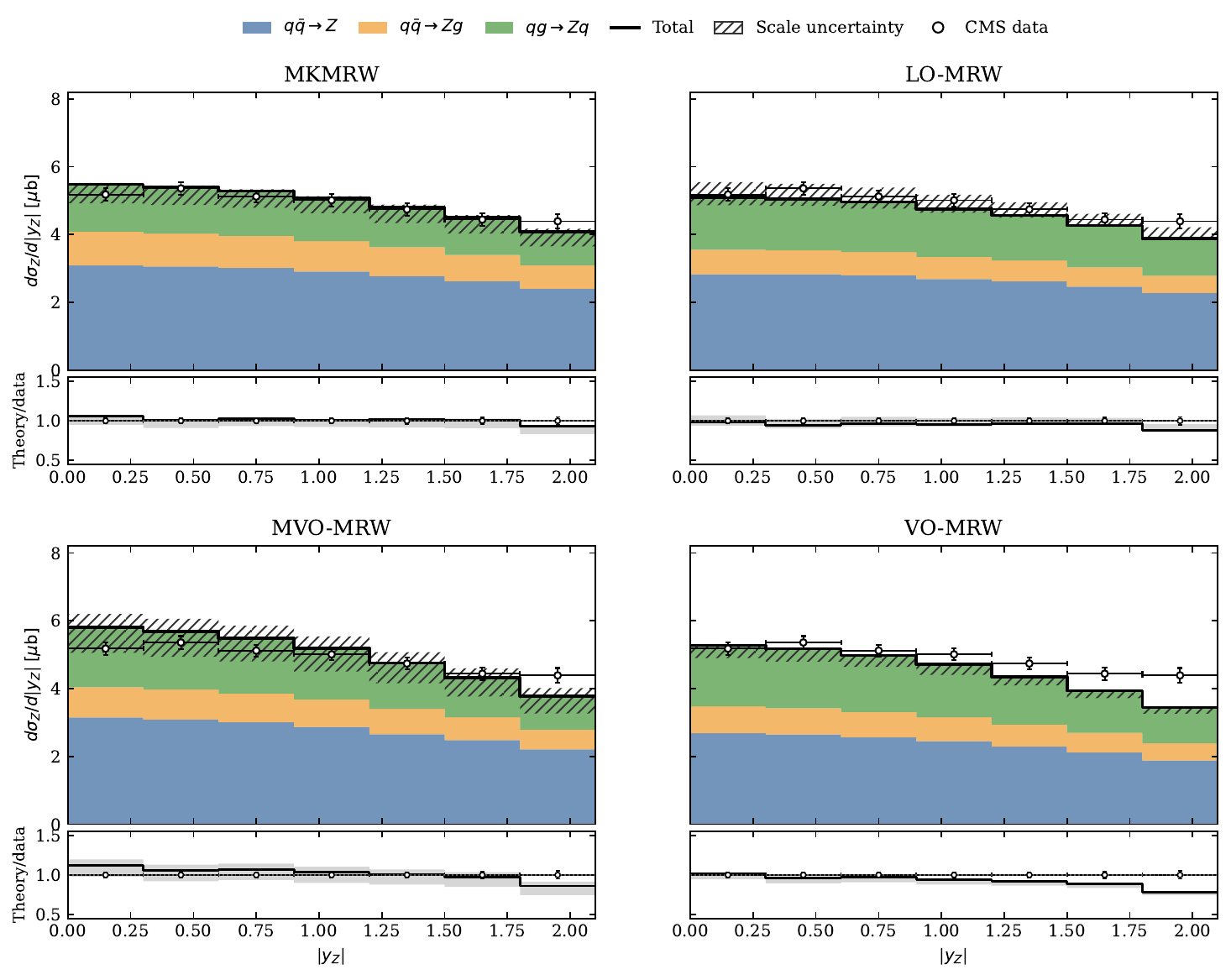}
	\caption{Subprocess contributions to the rapidity distribution for the four nuclear UPDF prescriptions. Colored regions show the additive central contributions of the Born $q\bar q$, real-emission $q\bar qg$, and $qg$ channels; the $qg$ contribution includes the corresponding crossings and antiquark channels. The black line denotes the total and the hatched band its scale uncertainty. Experimental error bars combine statistical and systematic uncertainties in quadrature. Lower panels show the total theory/data ratio and its scale band. The subprocess labels use $Z$ as shorthand for the full $Z\to\ell^+\ell^-$ contribution.}
	\label{fig:results-subprocess-y}
\end{figure*}

\begin{figure*}[tbp]
	\centering
	\includegraphics[width=0.98\textwidth]{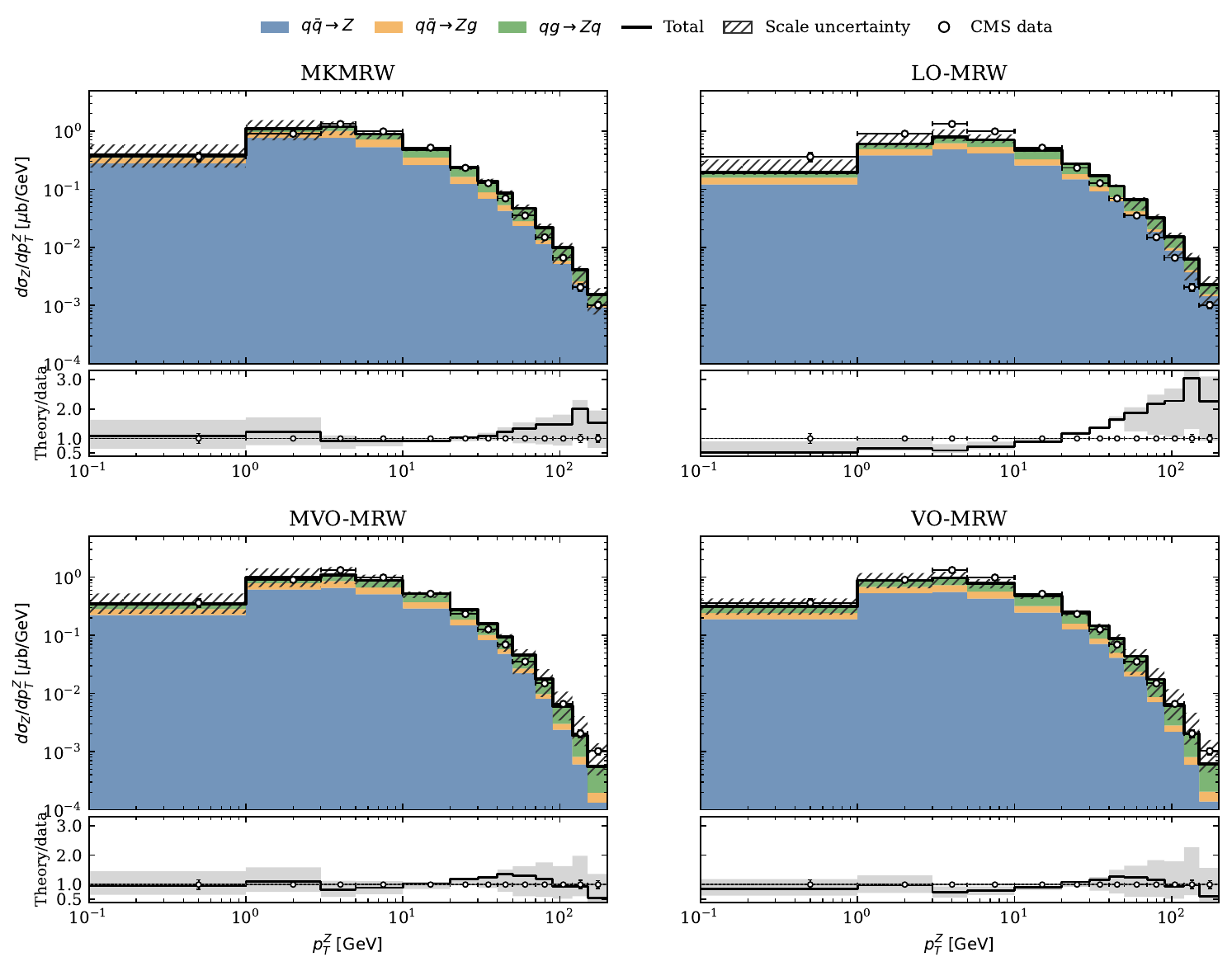}
	\caption{Subprocess contributions to the transverse-momentum distribution, with the same conventions as in Fig.~\ref{fig:results-subprocess-y}. The stacked regions are shown on a logarithmic vertical axis; their visual thickness is not proportional to their fraction of the total cross section.}
	\label{fig:results-subprocess-pt}
\end{figure*}

Figures~\ref{fig:results-subprocess-y} and~\ref{fig:results-subprocess-pt} show the contributions of the different partonic subprocesses to the rapidity and transverse-momentum distributions, respectively. In the rapidity distribution, the Born $q\bar q\to Z$ channel gives the largest individual contribution for all four UPDF prescriptions. Among the real-emission subprocesses, $qg\to Zq$ contributes more than $q\bar q\to Zg$. Together, these channels make a sizable contribution to the total cross section, although their relative contributions change only slightly across the rapidity range. This shows that the subprocess composition remains generally similar throughout the measured rapidity region.

The transverse-momentum distributions show a different behavior. At low $p_T^Z$, the Born channel dominates for all four models. As $p_T^Z$ increases, however, the relative contributions of the subprocesses become more dependent on the UPDF prescription. For MKMRW and LO-MRW, the Born channel remains the largest contribution even at high transverse momenta. In contrast, its contribution decreases more rapidly in VO-MRW and MVO-MRW, where the $qg\to Zq$ subprocess becomes dominant in the high-$p_T^Z$ region. This difference can be understood in part from the virtuality-ordering constraint, which restricts the transverse momenta of the incoming partons in all subprocesses. In the Born channel, the transverse momentum of the $Z$ boson is entirely determined by the incoming partons, $\boldsymbol{p}_T^Z=\boldsymbol{k}_{T1}+\boldsymbol{k}_{T2}$. For the real-emission subprocesses, however, the outgoing parton can balance a large $p_T^Z$, even when the incoming transverse momenta are relatively small. As a result, the virtuality-ordering constraint has a more direct effect on the Born contribution, while the real-emission channels remain important at high $p_T^Z$. The relative contributions also depend on the factorization scale and the phase-space separation prescription used to reduce double counting between subprocesses.

\section{Conclusions}
\label{sec:conclusion}
In this study, we investigated $Z$-boson production in Pb--Pb collisions at $\sqrt{s_{NN}}=5.02~\mathrm{TeV}$ using four nuclear UPDF prescriptions: LO-MRW, VO-MRW, MVO-MRW, and MKMRW. We calculated the rapidity and transverse-momentum distributions using the parton level KaTie event generator, with nCTEQ15 nuclear PDFs as the common collinear input. The predictions were compared with CMS measurements and the collinear MG5+EPPS16 calculation to examine the effects of different ordering constraints and normalization prescriptions.

We showed that all four models reproduce the general behavior of the measured rapidity distribution, with MKMRW giving the closest central prediction over most of the considered range. The differences become more apparent in the transverse-momentum distribution. LO-MRW underestimates the data at low $p_T^Z$ and gives a harder spectrum at high transverse momenta, while MKMRW improves the description in the low-$p_T^Z$ region. The virtuality-ordered models, VO-MRW and MVO-MRW, produce softer high-$p_T^Z$ spectra and agree more closely with the data in this region. We also found that normalization matching improves the VO-MRW prediction in some kinematic regions, but does not necessarily lead to better agreement over the entire distribution.

The subprocess contributions also differ between the models, particularly at high transverse momenta. While the Born channel remains the largest contribution for MKMRW and LO-MRW, the $qg\to Zq$ channel becomes dominant for the virtuality-ordered prescriptions in the high-$p_T^Z$ region. This behavior is partly related to the virtuality-ordering constraint, which limits the transverse momenta of the incoming partons. Although the scale uncertainties make it difficult to favor one model over the others, the comparison shows that the choice of ordering and normalization prescriptions can have a noticeable effect on the predicted distributions, even when the same nuclear PDFs are used.
	\section*{ACKNOWLEDGEMENTS}
	We would like to express our gratitude to the Institute for Research in Fundamental Sciences (IPM) for their financial support and services.
	
	\bibliography{article}
\end{document}